\documentclass{iau}

\usepackage{graphicx}

\title[Crust Failure and Plastic Flow] 
{Magnetic Field Evolution in the Crust of Neutron Stars: Crust Failure and Plastic Flow}

\author[Konstantinos N. Gourgouliatos]   
{Konstantinos N. Gourgouliatos$^1$}

\affiliation{$^1$University of Patras, Department of Physics, Patras, 26504, Greece \\email: {\tt kngourg@upatras.gr}}

\pubyear{2022}
\volume{363}  
\jname{Neutron Star Astrophysics at the Crossroads: Magnetars and the Multimessenger Revolution}
\editors{Prof. E. Troja \& Prof M. Baring, eds.}
\begin{document}

\maketitle

\begin{abstract}
The evolution of the magnetic field in neutron star crusts because of the Hall effect has received significant attention over the last two decades, which is strongly justified because of the dominance of this effect in highly magnetised neutron stars. However, the applicability of the Hall effect is based on the assumption that the crust does not fail and sustains its rigidity. This assumption can be violated for substantially strong magnetic fields. If this is the case, the evolution of the magnetic field is described by a different set of equations, which include the effects of a non-rigid crust. In this talk, after a brief review of the main characteristic of the Hall evolution, I will discuss the impact a plastic flow of the crust has on the magnetic field, studying axisymmetric models. Moreover, the way the crust fails impacts the overall evolution, with major differences appearing if the failure is local, intermediate or global. Quite remarkably, crustal failure and plasticity do not annul the Hall effect, and under certain circumstances they may even lead to a more dramatic evolution. I will discuss the impact of these effects in the context of neutron star timing behaviour, with special focus on timing noise, outbursts and glitches. 
\keywords{stars: neutron, (magnetohydrodynamics:) MHD, magnetic fields.}
\end{abstract}

\firstsection 
\section{Introduction}

The magnetic field plays a central role in neutron stars being one of the key factors determining their observable behaviour. Magnetars are strongly magnetised neutron stars manifesting their presence through X-rays and soft gamma rays (\cite{2017ARA&A..55..261K,2015RPPh...78k6901T}). They radiate as persistent sources or through explosive events in the form of bursts, outbursts and flares (\cite{2005Natur.434.1107P, 2011ASSP...21..247R,2018MNRAS.474..961C}), and it is usual for magnetars to switch from quiescent behaviour to outbursts. Their magnetic field is key to both their persistent emission and the explosive events. The former has been attributed to the decay of the magnetic field and the released of Ohmic heating (\cite{2007A&A...470..303P}) or the acceleration of particles in the magnetosphere which bombard the surface of the neutron star and heat it (\cite{2009ApJ...703.1044B}). The explosive events are related to sudden changes of the magnetic field, either through twisting and rapid untwisting of the magnetospheric field (\cite{1995MNRAS.275..255T,2018ASSL..457...57G}) or the evolution of the internal magnetic field which leads to crustal failure (\cite{2011ApJ...741..123P,2011ApJ...727L..51P,2015MNRAS.449.2047L,2016ApJ...824L..21L,2017ApJ...841...54T}). Given the wide variety in the properties of the explosive events it is possible that distinct mechanisms are related to flares and bursts, involving different parts of the star. 

As the magnetic field is believed to drive these processes, its modelling has received significant attention. The main mechanisms of evolution are the Hall effect, Ohmic dissipation and ambipolar diffusion, which have been studied using realistic microphysics for the crust and have been related to observable properties (\cite{Goldreich:1992}). The evolution under the Hall effect is based on the assumption that the crust is strong enough to maintain dynamical equilibrium, through elastic forces. Thus, the momentum equation is satisfied identically. While this condition holds for weak and moderately strong magnetic fields, a sufficiently strong magnetic field can generate stresses which cannot be balanced by the crust's elastic forces. Thus, from a theoretical perspective one expects that there is a limitation on the strength of the magnetic field that can evolve due the Hall effect. From the observational side, the variety of explosive events is a possible indication of crust failure. Several mechanisms have been proposed for the explanation of events and a few of them involve some form of crustal yielding. Therefore, both from an observational and a theoretical perspective crust failure is a likely scenario. Crust failure does not imply that the field halts its evolution, however post failure evolution cannot be determined entirely by the Hall and Ohmic effect, but it will also depend on the state of the failed crust. Moreover, it is important to clarify under what conditions the crust fails, and once this happens what part of the crust becomes affected. Thus, the failed crust can continue its evolution through a plastic flow given that the magnetic field stresses are not completely relaxed. The properties of this flow will be given by the solution of the appropriate Navier-Stokes equation, where the crust's plastic viscosity plays a key role.

The structure of this paper is as follows: in section \ref{sec:formulation} we present the physical conditions that lead to plastic flow and the equations that are solved for. In section \ref{sec:simulations} we present the simulation developed for this problem and the results found. We discuss these results in the context of neutron star physics and we conclude in section \ref{sec:discussion}. 

\section{Plastic Flow Formulation}
\label{sec:formulation}

Following the formation of a neutron star, its crust freezes into a solid form, comprising a body centred cubic Coulomb lattice (\cite{2008LRR....11...10C}). This freezing process involves also the magnetic field, which relaxes within a few Alfv\'en time-scales to a magnetohydrodynamical equilibrium (\cite{2004Natur.431..819B}). The magnetic field and the recently frozen crust start in a state of dynamical equilibrium. This does not imply that the field itself has zero Maxwell stresses, but rather the structure of the crust is such that these stresses are balanced. Moreover, despite the field being in a magnetohydrodynamical equilibrium, it may start immediately evolving because of the Hall effect, as the structure of the frozen crust is different from that of a fluid proto-neutron star. In general an MHD equilibrium is not identical to a Hall equilibrium (\cite{2013MNRAS.434.2480G}). As the magnetic field evolves due to the Hall effect it changes its structure, and the Maxwell stresses it exerts onto the crust evolve with time as well. The Maxwell stresses of the crust are given by the following expression:
\begin{eqnarray}
{\cal M}_{ij}=\frac{1}{4\pi }B_i B_j \,,
\end{eqnarray}
where $B_{i}$ are the components of the magnetic field. The off-diagonal elements of the Maxwell stress tensor correspond to shear terms, while the diagonal element to pressure terms. Given the high pressure of the crust, its failure will be primarily due to the shear terms, rather than compression, thus, for the consideration of failure we need to account only for the off-diagonal terms:
\begin{eqnarray}
\tilde{\cal M}_{ij}=\frac{1}{4\pi }\left(B_i B_j -\frac{1}{3}B^2\delta_{ij}\right)\,.
\end{eqnarray}
 These stresses may become comparable to the yield stress of the crust, which is also the elastic limit. The yield stress of crust depends on the microphysical properties of the crust, and is given by the following expression (\cite{2010MNRAS.407L..54C}): 
\begin{eqnarray}
\tau_{el}=\left(0.0195-\frac{1.27}{\Gamma-71}\right)\frac{Z^2 e^2 n_{\rm I}}{a_{\rm I}}\,,
\end{eqnarray}
where $Z$ is the atomic number, $e$ is the electron charge, $\Gamma$ the Coulomb parameter, $n_{\rm I}$ the ion number density, and $a_{\rm I}$ the inter-ion separation. The yield stress is a strong function of the neutron star density and therefore the depth.

We then apply the von Mises criterion for the failure conditions of the crust to determine whether crust fails (\cite{2019MNRAS.486.4130L}):
\begin{eqnarray}
\tau_{el} \leq \sqrt{ \frac{1}{2}\left( \tilde{\cal M}_{ij}^0-\tilde{\cal M}_{ij} \right) \left( \tilde{\cal M}_{ij}^0-\tilde{\cal M}_{ij} \right) }\,,
\end{eqnarray}
which is can be expressed in terms of the magnetic field as follows:
\begin{eqnarray}
\tau_{el}\leq \frac{1}{4\pi} \sqrt{\frac{1}{3}B_0^4+\frac{1}{3}B^4+\frac{1}{3}B_0^2B^2-\left(\vec{B}\cdot \vec{B}_0\right)^2}\,,
\label{tau_el}
\end{eqnarray}
where $\vec{B}$ is the  present magnetic field and $\vec{B}_0$ is the initial magnetic field. Thus, if the magnetic field has undergone a substantial change, it may produce a significant plastic flow. We note that, this implies that even the decay of the magnetic field can lead to failure. This is because the crust has frozen into a state that maintains equilibrium for given Maxwell stresses, and if the latter are removed the crust goes off equilibrium. The evolution of the failed crust can be then approximated by a plastic Stoke's flow, as the velocities are relatively  low, thus acceleration and quadratic terms can be neglected. Thus, the plastic flow velocity satisfies the equation (\cite{2016ApJ...824L..21L}):
\begin{eqnarray}
4\pi \nu \nabla^2 \vec{ v}_{pl}=-4\pi \vec{\nabla}\cdot\left(\tilde{\cal M}-\tilde{\cal M}_0\right)\,.
\label{EQ_STRESS}
\end{eqnarray}
The solution of the above equation for given Maxwell stress provides the plastic flow velocity. This velocity can be included into the Hall-induction equation, which is then integrated and gives the magnetic field evolution (\cite{2016ApJ...824L..21L}):
\begin{eqnarray}
\partial_{t} {\vec B} = -{\vec \nabla}\times \left[\left(\frac{c}{4 \pi e n_e}{\vec \nabla} \times {\vec B}-\vec{v}_{pl}\right)\times {\vec B}  +\frac{c^2}{4 \pi \sigma}{\vec \nabla} \times {\vec B}\right]\,.
\label{PLASTIC_EQ}
\end{eqnarray}

\section{Numerical Simulations}
\label{sec:simulations}
 \begin{figure}[b]
\begin{center}
 \includegraphics[width=0.9\textwidth]{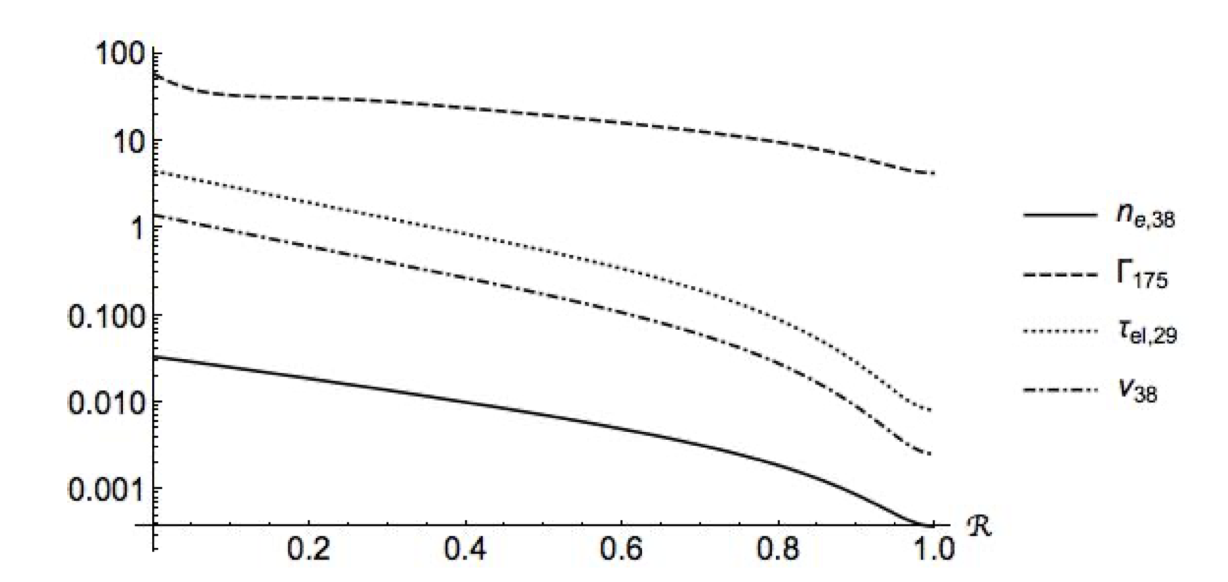} 
 \caption{The values of the microphysical parameters as a function of the distance from the base of the crust, used in the numerical simulations: $n_{e,38}$ is the electron number density in units of $10^{38}$cm$^{-3}$, $\Gamma_{175} $ is the Coulomb parameter in units of $175$, $\tau_{el,29}$ is the critical stress in units of $10^{29}$g~cm$^{-1}$~s$^{-2}$ and $\nu_{38}$ is the plastic flow viscosity in units of $10^{38}$g~cm$^{-1}$~s$^{-1}$. The horizontal axis is the distance from the base of the crust and ${\cal R}=1$ corresponds to the neutron drip point. Figure reproduced from \cite{2019MNRAS.486.4130L}.}
   \label{Fig:1}
\end{center}
\end{figure}

The complexity of the system of equations describing the a plastic flow in the crust and the subsequent magnetic field evolution allows only very limited analytical solutions (\cite{2016ApJ...824L..21L}). Therefore a numerical implementation is required. We have explored two main geometries, a cartesian box (\cite{2019MNRAS.486.4130L}) and an axisymmetric spherical shell (\cite{2021MNRAS.506.3578G}). In both geometries we do not allow  the plastic flow to have components in the vertical or radial direction assuming that the crust is stably stratified. We have further assumed that the crust is incompressible, thus the divergence of the plastic flow velocity is zero. Regarding the microphysical parameters, we have used the functions described in the above papers, and summarised in the Figure (\ref{Fig:1}). This range of parameters correspond to a neutron star 1.4 solar masses whose equation of state is based on the SLy4 interaction (\cite{2001A&A...380..151D}).

 \subsection{Cartesian Simulations}
 \begin{figure}
\begin{center}
 \includegraphics[width=0.9\textwidth]{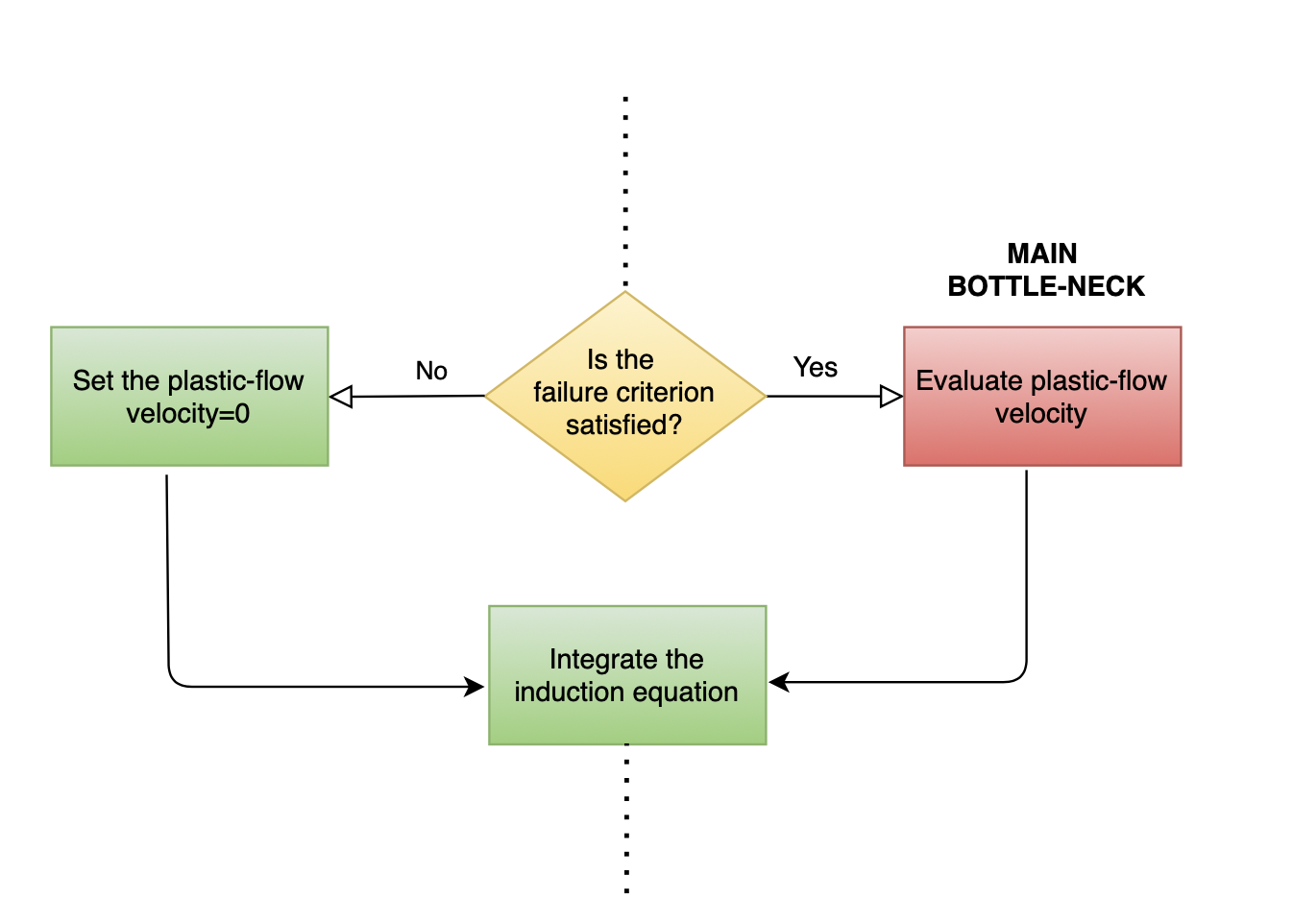} 
 \caption{Outline of the numerical procedure followed for the evaluation of the failure criterion and the numerical integration of the induction equation.}
   \label{Fig:2}
\end{center}
\end{figure}

 In the cartesian geometry  we have studied a square domain, where have assumed plane-parallel symmetry: the magnetic field has all three components but it varies only along two out of three coordinates. The square slab has a side whose length is equal to the thickness of the crust from the crust-core boundary to the neutron drip point (0.5 km). We adopted the numerical procedure outlined in Figure \ref{Fig:2}. A set of initial conditions for the magnetic field structure is chosen, which is then integrated in time using equation (\ref{PLASTIC_EQ}), with zero initial plastic flow velocity. Then, we check whether the failure criterion of equation (\ref{tau_el}) is satisfied. If this condition is met anywhere within the domain, equation (\ref{EQ_STRESS}) is solved for the plastic flow velocity which is then substituted in (\ref{PLASTIC_EQ}). If the failure condition is not satisfied anywhere in the crust then the usual Hall-MHD equation is integration, which is in effect equation (\ref{PLASTIC_EQ}) with zero plastic flow velocity. The boundary conditions at the base of the crust have a zero magnetic field; at the exterior the field is matched to a current-free field; and we have adopted periodic boundary conditions at the sides. We have utilised a modified version of codes employed for the Hall effect previously (\cite{2015MNRAS.453L..93G,2016MNRAS.463.3381G}). 
 
 We have explored a wide variety of initial conditions regarding the structure of the magnetic field. As the plastic flow viscosity is a parameter that cannot be accurately determined by the microphysics we have explored a range of values, a factor of $10$ higher and a factor of $10$ lower than the basic one shown in Figure \ref{Fig:1}. Moreover, we have run simulations where only the Hall effect is included to compare the impact of plastic flow.
 
\begin{figure}
\begin{center}
 (a) \includegraphics[width=0.45\textwidth]{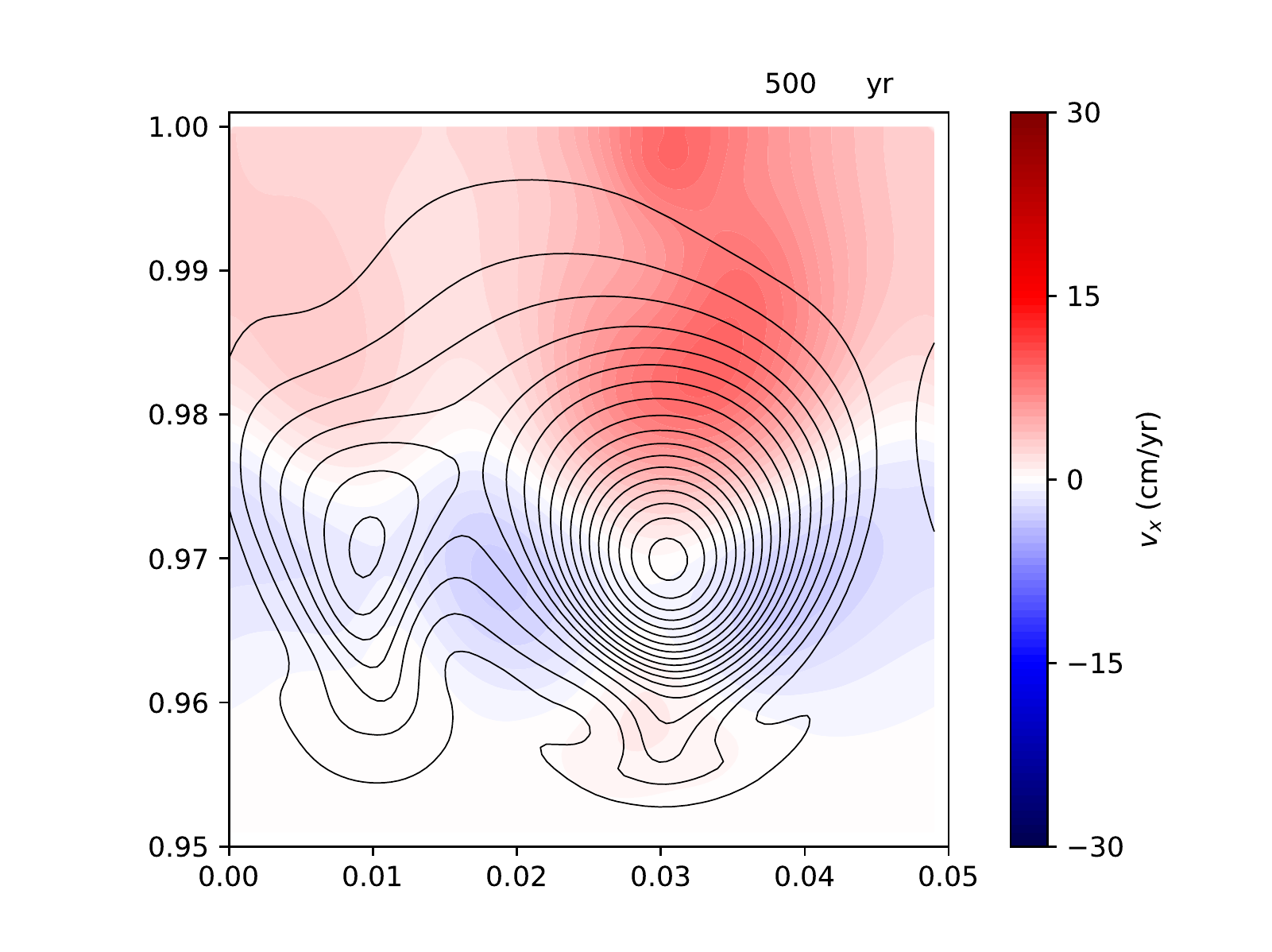} 
 (b) \includegraphics[width=0.45\textwidth]{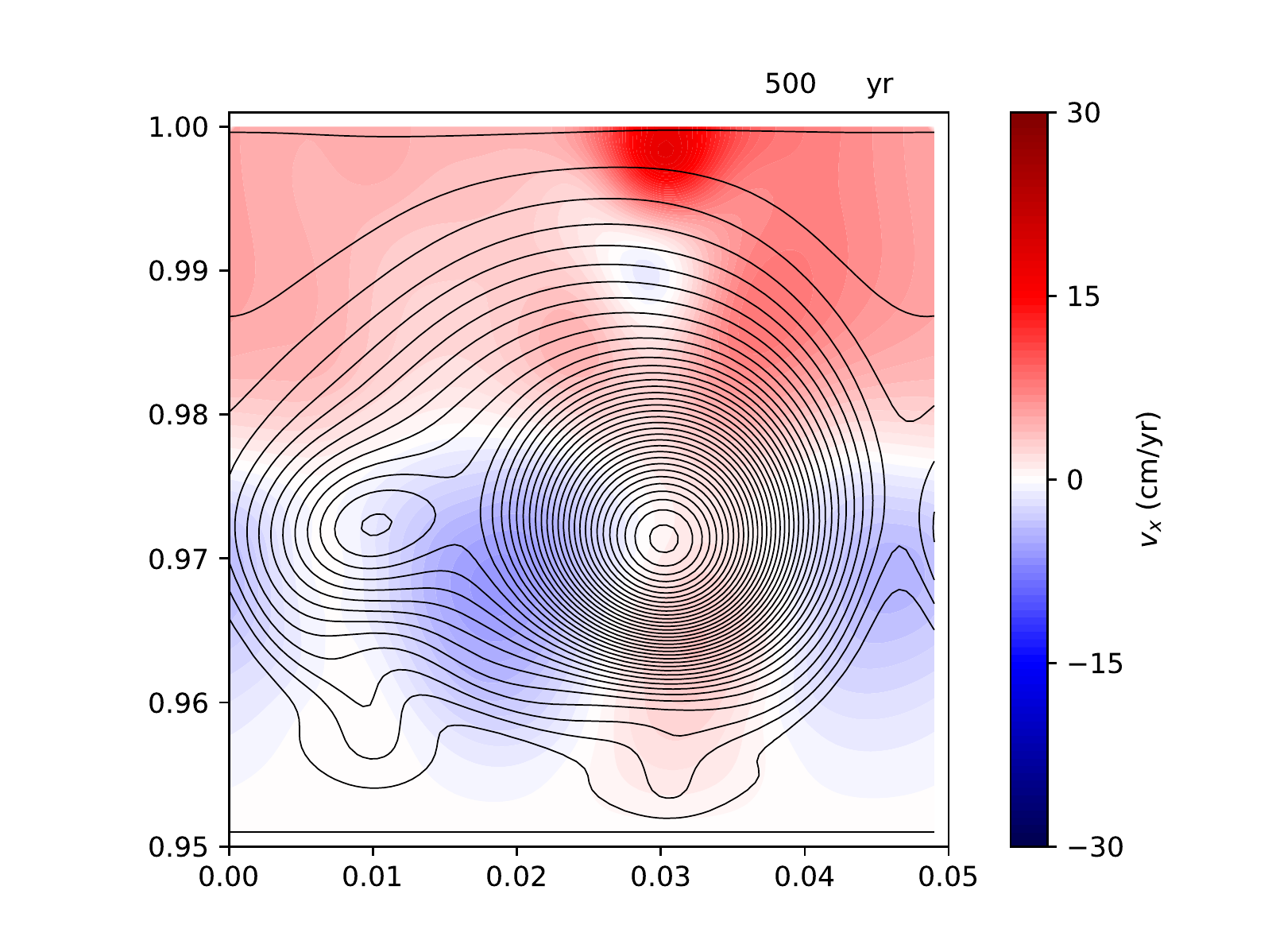} 
  (c)  \includegraphics[width=0.45\textwidth]{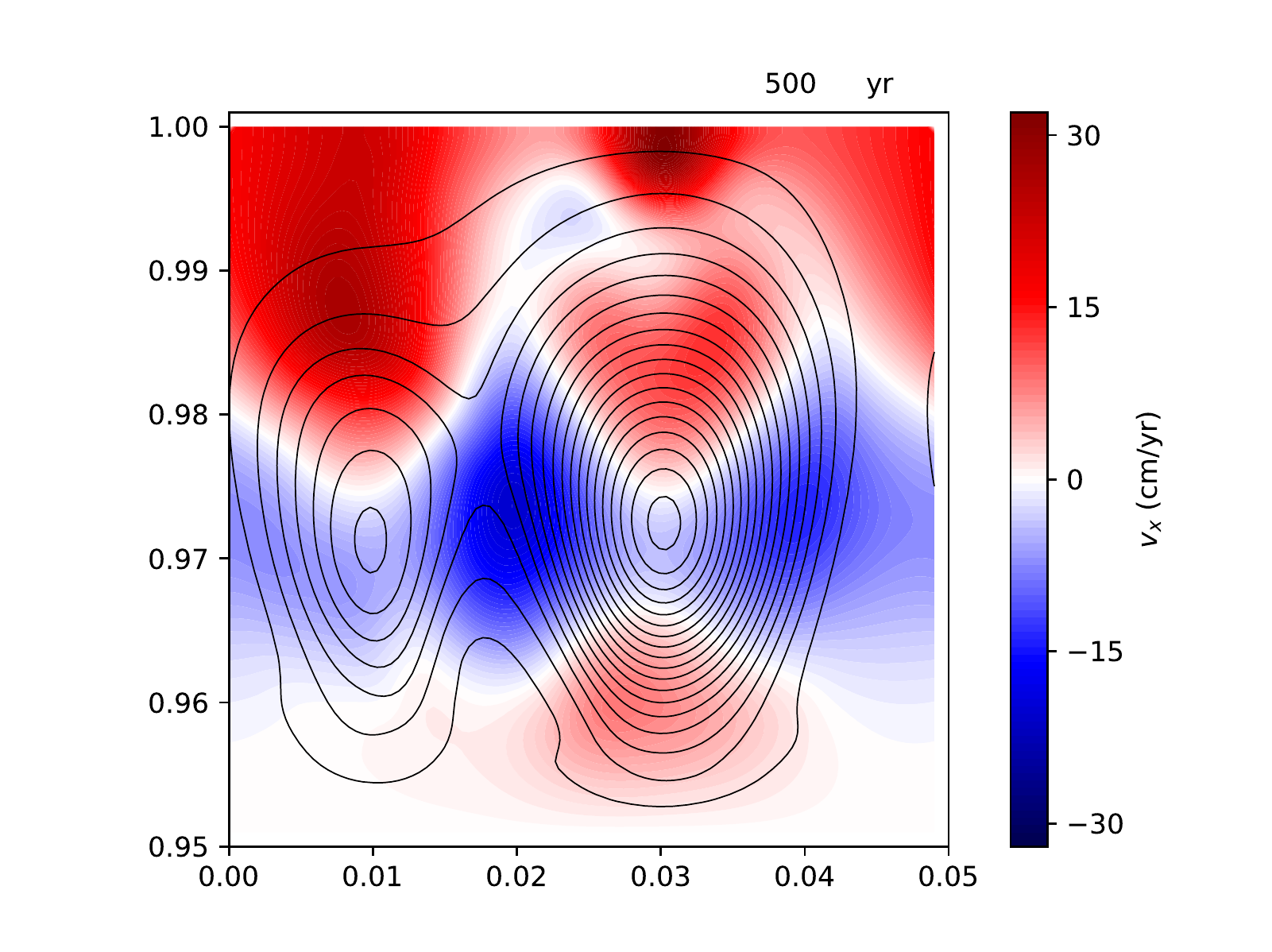} 
  (d)  \includegraphics[width=0.45\textwidth]{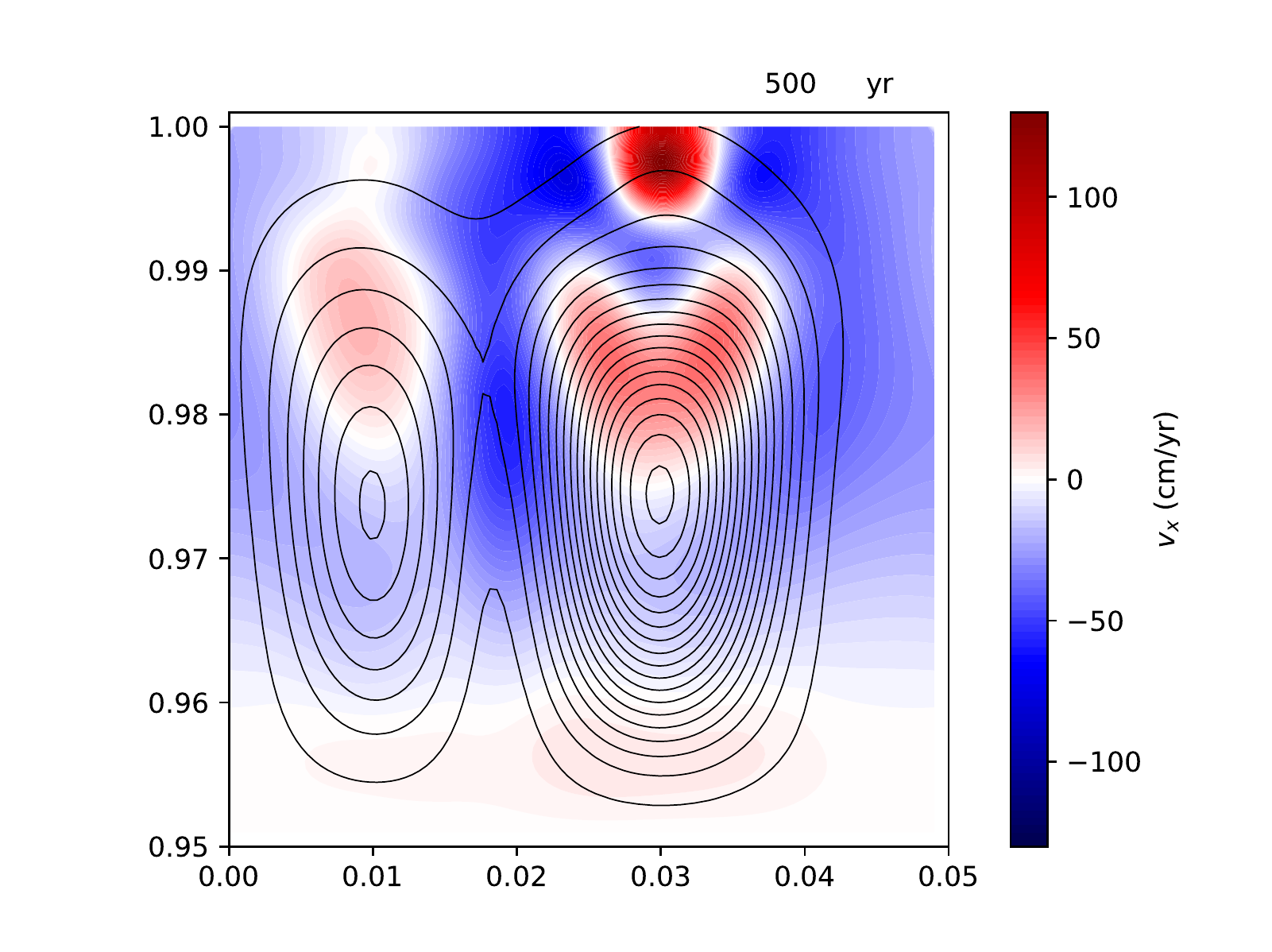} 
 \caption{ Plastic flow velocity at 500 years of neutron star age for the cartesian simulation shown in colour with poloidal field lines shown in black. The lengths given are in units of the neutron star radius. The vertical dimension corresponds to the depth of the crust, with $0.95$ being the crust-core interface and $1$ the neutron drip point where the surface of the star is set. (a) The maximum initial magnetic field strength is $2\times 10^{15}$G and the plastic flow viscosity at the base of the crust is $10^{38}$ g~cm$^{-1}$~s$^{-1}$. (b) The maximum initial magnetic field strength is $4\times 10^{15}$G and the plastic flow viscosity at the base of the crust is $10^{38}$ g~cm$^{-1}$~s$^{-1}$. (c) The maximum initial magnetic field strength is $2\times 10^{15}$G and the plastic flow viscosity at the base of the crust is $10^{37}$ g~cm$^{-1}$~s$^{-1}$.  (d) The maximum initial magnetic field strength is $2\times 10^{15}$G and the plastic flow viscosity at the base of the crust is $10^{36}$ g~cm$^{-1}$~s$^{-1}$. Figure reproduced from Lander and Gourgouliatos (2019). }
   \label{Fig:3}
\end{center}
\end{figure}

We have run the simulation for $2\times 10^{3}$ years of neutron star age, until the initial magnetic field evolution saturates. The crust fails within the first $100$ years of the evolution if the maximum strength of the magnetic field exceeds $10^{15}$G and develops a plastic flow velocity whose maximum value depends on the plastic flow viscosity chosen and can reach up two $130$ cm/yr, Figure \ref{Fig:3}. 

 \begin{figure}
\begin{center}
 (a) \includegraphics[width=0.4\textwidth]{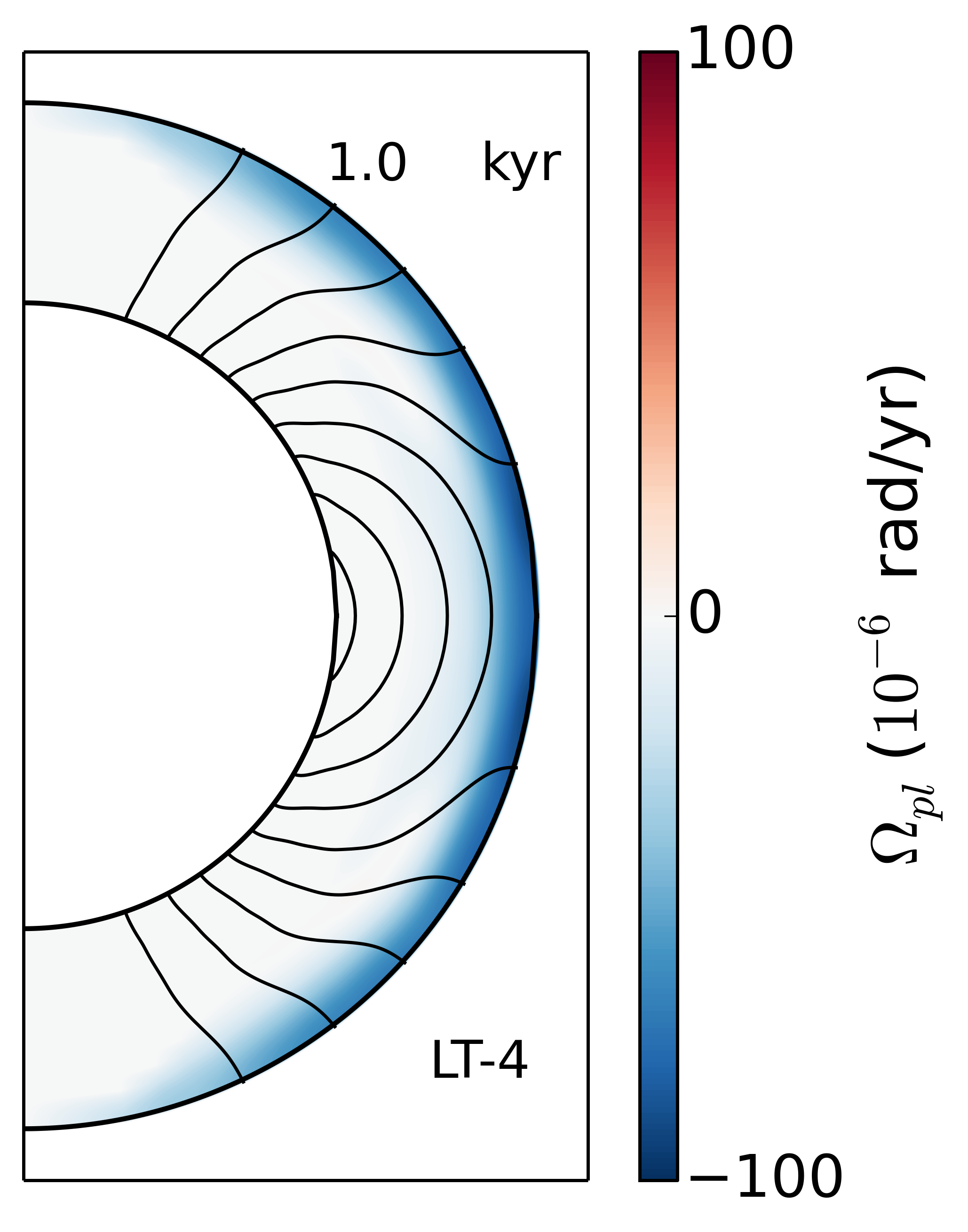} 
 (b) \includegraphics[width=0.4\textwidth]{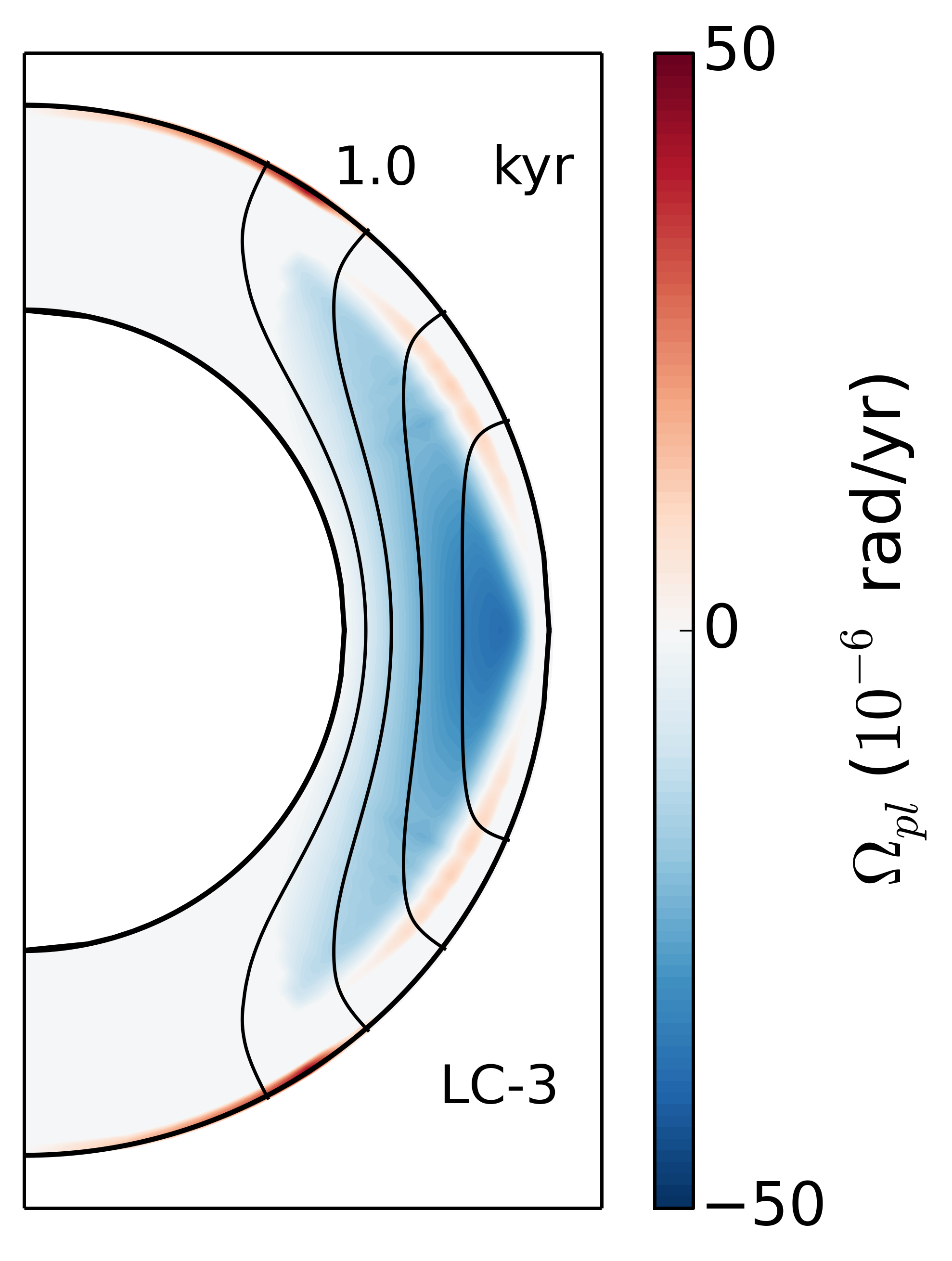} 
  (c)  \includegraphics[width=0.4\textwidth]{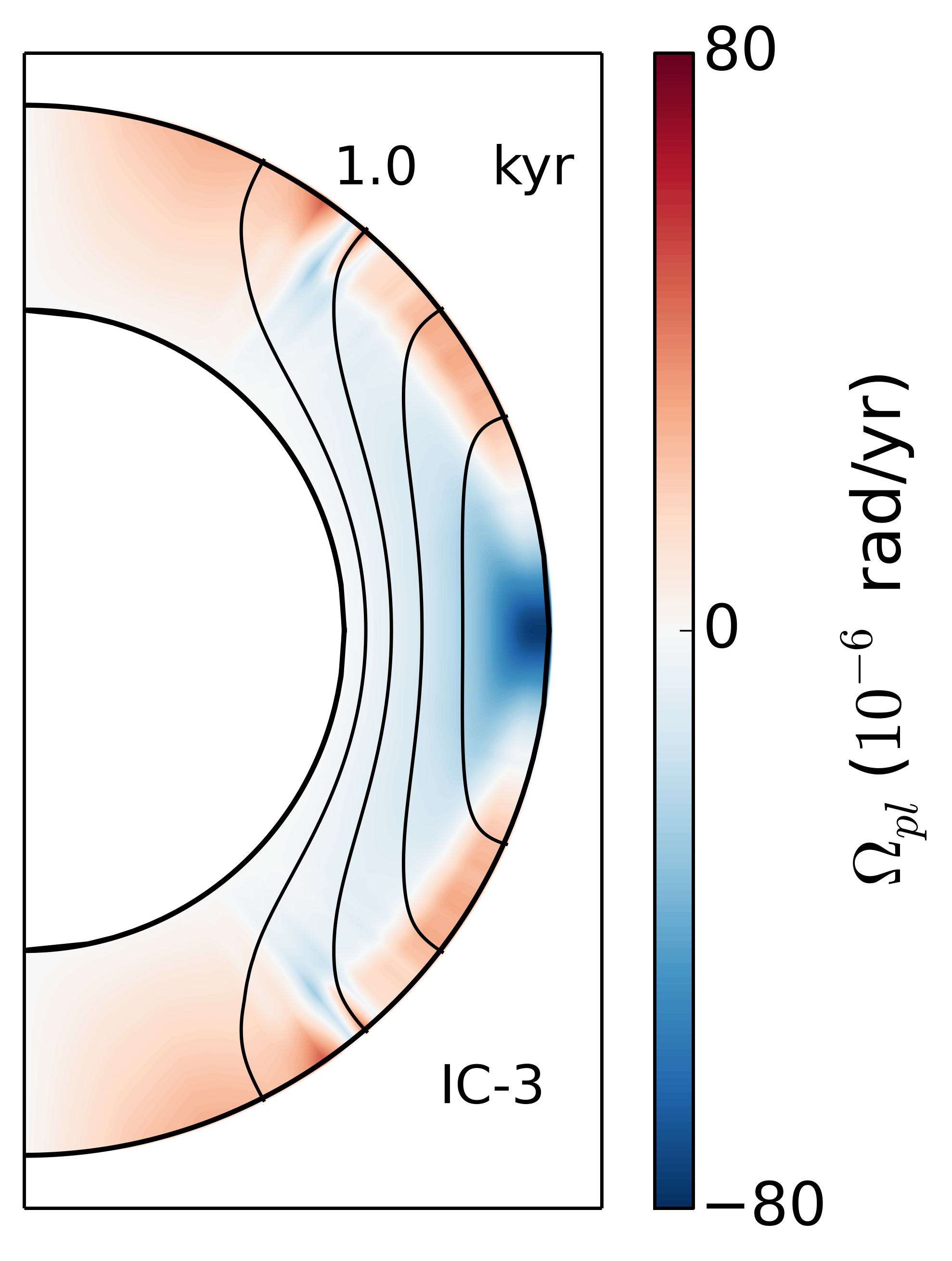} 
  (d)  \includegraphics[width=0.4\textwidth]{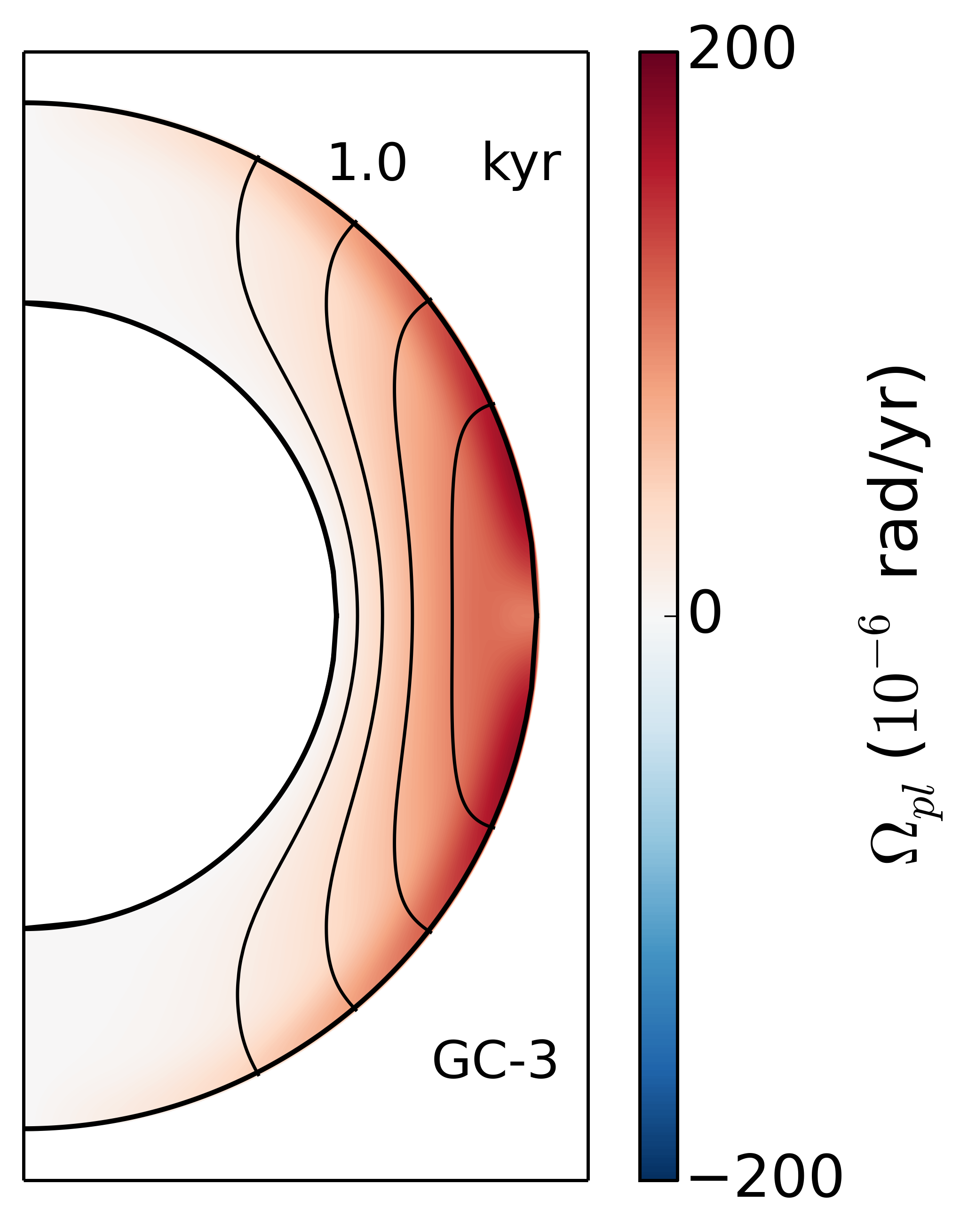} 
 \caption{ Plastic flow angular velocity at 1000 years of neutron star age for the axisymmetric simulation shown in colour, with black lines indicating the poloidal field line structure, for different types of failure and initial magnetic field structure. In all panels, the initial dipole magnetic field strength is $10^{14}$G and the plastic flow viscosity at the base of the crust is $10^{38}$ g~cm$^{-1}$~s$^{-1}$. The crust thickness corresponds  to $0.05$ of the neutron star's radius and has been magnified by a factor of 8 for visualisation purposes. (a) Local failure criterion with field lines penetrating the core. (b) Local failure criterion with field lines contained in the crust. (c) Intermediate failure criterion with field lines contained in the crust. (d) Global failure criterion with field lines contained in the crust. Figure reproduced from Gourgouliatos and Lander (2021).}
   \label{Fig:4}
\end{center}
\end{figure}
 \begin{figure}
\begin{center}
 (a) \includegraphics[width=0.4\textwidth]{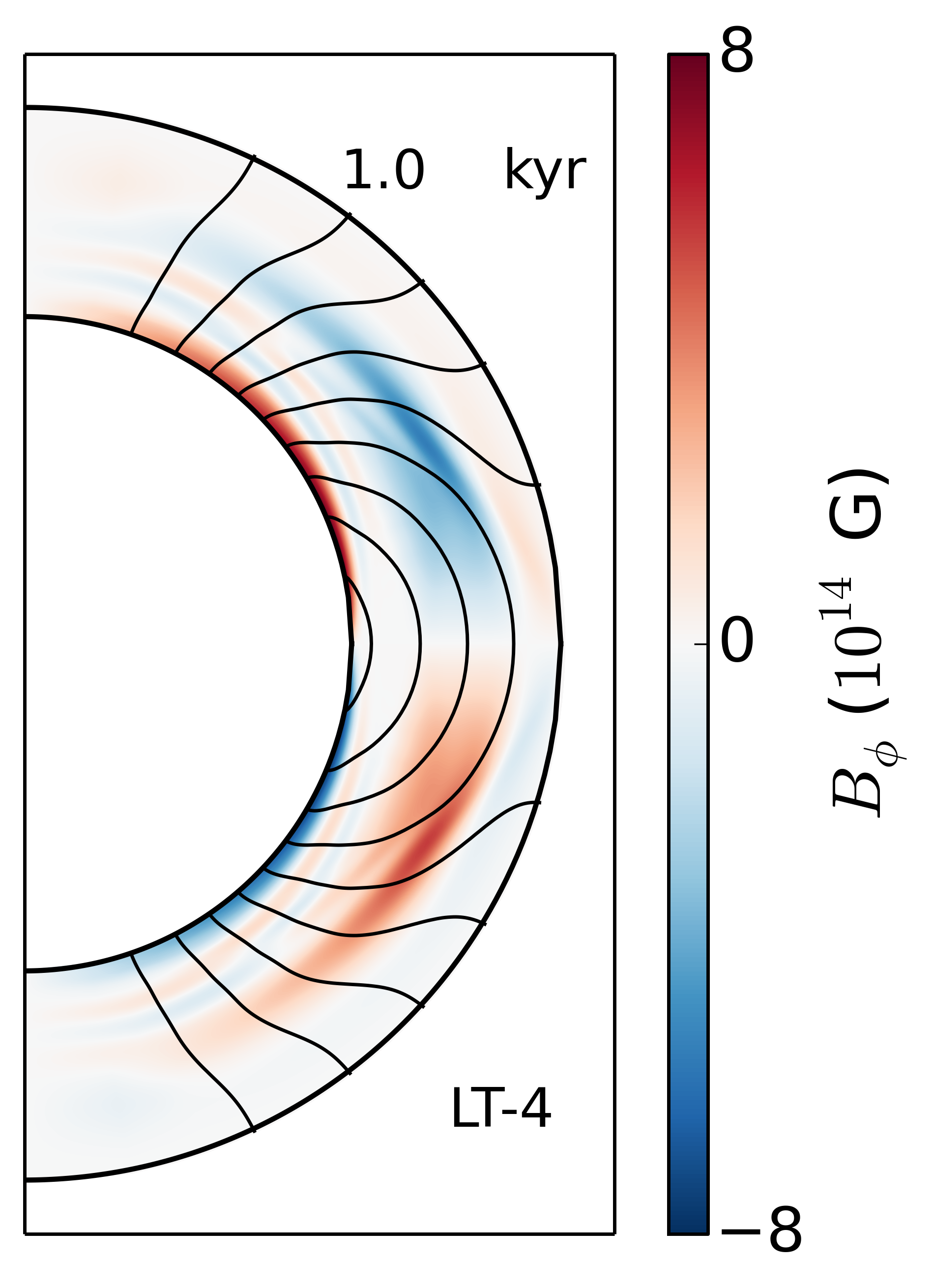} 
 (b) \includegraphics[width=0.4\textwidth]{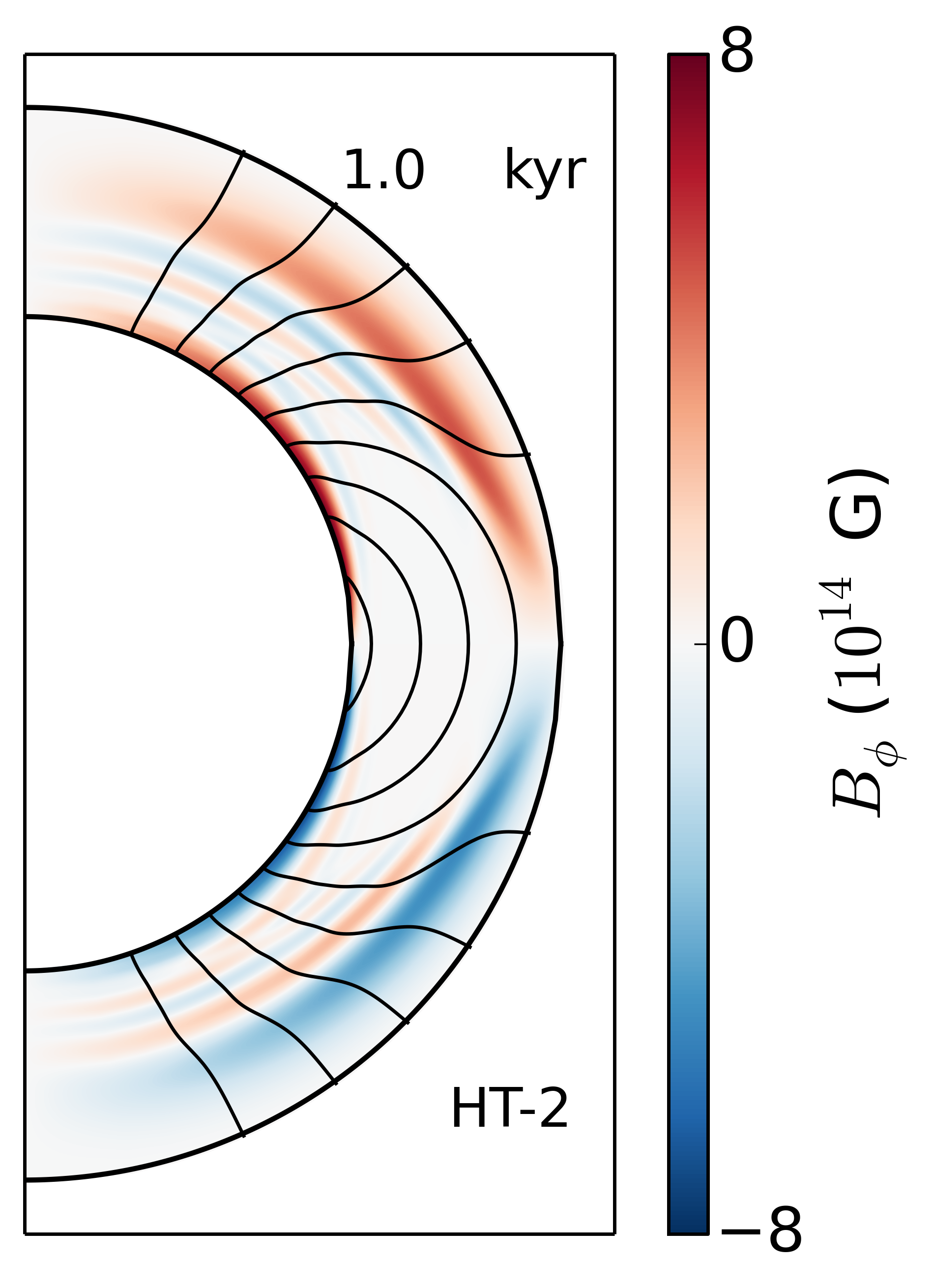} 
 \caption{ The toroidal field at 1000 years of neutron star age for the axisymmetric simulation shown in colour, with black lines indicating the poloidal field line structure, for two simulations with identical initial conditions and a dipole initial magnetic field strength $10^{14}$G. The crust thickness corresponds  to $0.05$ of the neutron star's radius and has been magnified by a factor of 8 for visualisation purposes. The simulation shown in plot (a) has a plastic flow, utilising a local failure criterion and the one shown in plot (b) has only evolution under the Hall effect. The plastic flow viscosity at the base of the crust for both simulations shown in plot is $10^{38}$ g~cm$^{-1}$~s$^{-1}$. Figure reproduced from Gourgouliatos and Lander (2021).}
   \label{Fig:5}
\end{center}
\end{figure}

\subsection{Axisymmetric Simulations}

We have also explored axisymmetric simulations of a spherical shell representing the crust. Unlike the cartesian box simulations that are local, these simulations are global. Even though equation (\ref{tau_el}) provides a robust criterion for failure, the part of the star that is affected by such failures needs some further consideration. Clearly the parts of the crust where this condition holds will be affected and undergo a plastic flow, but it is likely that regions in the vicinity of the failed part of the crust will be affected as well. Given this uncertainty we have employed three different scenarios. First, a local failure scenario where there is a plastic flow only in the parts of the crust where the von Mises criterion is satisfied. Then, an intermediate failure scenario, where a plastic flow is allowed everywhere in the crust, but the source term of equation (\ref{EQ_STRESS}) is non-zero only in the parts of the crust where the von Mises criterion is satisfied. Finally, a global failure scenario, where both the flow and the source term of equation (\ref{EQ_STRESS}) are non-zero if the von Mises criterion is satisfied even in a single point in the crust. We remark that the global failure scenario is primarily viewed as an extreme case for the quantification of the maximal effects of plastic flows. Regarding the boundary conditions we have assumed a vacuum field at the exterior of the star; on the axis we use reflecting boundary conditions; and at the base of the crust we keep the radial field constant and set the derivative of the toroidal field equal to zero. Regarding the microphysics we utilise the same profile as in the cartesian simulations. 

The numerical implementation of this problem is based in a suitable adaptation of the numerical codes used previously for problems related to the Hall effect (\cite{2014MNRAS.438.1618G}), where an additional module for the plastic flow has been included. As in the cartesian case, we have explored a broad array of initial conditions for the structure and the strength of the magnetic field, as well as the value of the plastic viscosity, for all three types of failure. We have also studied cases where the evolution is only due to the Hall effect. We have run the simulations for $10^{4}$ years of neutron star life to focus on the effects we expect to operate in magnetars and young neutron stars. We have monitored the magnetic field evolution, the plastic flow velocity, the electron fluid velocity and the magnetic stress, with some indicative results shown in Figure \ref{Fig:4} and \ref{Fig:5}.

\section{Discussion}
\label{sec:discussion}

This wide exploration of the effect of plastic flow in the crusts of neutron stars has demonstrated an additional evolutionary channel for the magnetic field beyond the Hall effect that has been central in such studies so far (\cite{sym14010130}). The impact of plastic flow has started to receive significant attention recently focusing on applications such as the role global flows within the star, the limits of crust elasticity and the evolution of crustal fractures (\cite{2019MNRAS.488.5887S,2020MNRAS.491.1064G,2020MNRAS.494.3790K,2021MNRAS.502.2097K,2021MNRAS.506.2985K,2021MNRAS.506.3936K,2022arXiv220101881K}).
The study of plane-parallel and axisymmetric models, with realistic microphysical parameters allow us to draw the following conclusions. First, the strength of the magnetic field is crucial for the initial of plastic flow. In general magnetic fields below the $10^{14}$ G are too weak to lead to failure and initiate a plastic flow. Given that the breaking stress is a function of radius, a moderately strong field of $\sim 10^{14}$ G near the surface may lead to failure, whereas a magnetic field of $\sim 10^{15}$G at the base of the crust will be too weak to break it. Second, while in general the plastic flow opposes the electron flow associated to the Hall effect, it does not completely annul it. Simulations where the plastic flow is in the entire domain, such as cartesian simulations where only a slab of the crust is solved for and the spherical shell ones with global failure, stop the Hall evolution, while at the same time develop rather fast plastic flow velocities. If a local or intermediate type of failure is postulated, the evolution becomes more complex, as large parts of the field evolve only because of the Hall effect while other regions undergo  plastic flow. This may lead to regions where the magnetic field is even more twisted than it would have been if it was evolving only due to the Hall effect, effectively switching the polarity of the toroidal field in parts of the star, as the plastic flow velocity overreacts to the Hall effects locally and twists it in the opposite direction, as shown in Figure \ref{Fig:5}. Third, the plastic flow partially releases the stresses developed in the crust, but they are not completely relaxed. This is because the plastic flow is rather constrained due to the crust geometry, the symmetries and the assumption of stable stratification. The latter prevents any radial motion, therefore, such stress components cannot be relaxed by a plastic flow.  

Plastic flows appearing in neutron star bursts and outbursts may lead to drifting of the emitting region, if the patch of the crust where the burst is localised flows and does not remain stationary on the star's surface. This may be related to gradual phase shifts observed in some outbursts (\cite{2022ApJ...924L..27Y}). Moreover, the viscous nature of plastic flows contributes to some extra dissipation, which can further contribute to the magnetothermal evolution models that have successfully modelled several properties of the neutron star population (\cite{2013MNRAS.434..123V,2020ApJ...903...40D, 2021NatAs...5..145I}). We note however, that the fact that plastic flow in general opposes the Hall effect may lead to a decrease of the rate of dissipation, which is accelerated by the impact of the Hall effect. 

The rotational evolution of neutron stars also exhibits some irregularities either in the form of timing noise, or sudden changes of their rotational frequency in the form of glitches (\cite{2008ApJ...673.1044D}), and the even more puzzling spin-down events known as anti-glitches (\cite{2013Natur.497..591A}). Timing noise is higher in strongly magnetised neutrons stars (\cite{2013ApJ...773L..17T}). A plastic flow affecting the inner parts of the crust may impact the pinning of the superfluid vortices residing in the inner crust and the outer core and thus trigger glitches, and it may even generate regions flowing in the opposite direction of the neutron star rotation, thus create pockets of negative angular momentum. The sudden exchange of this angular momentum may lead to sudden spin-down events. 

We remark however, that the current level of modelling is constrained to simple geometries, which substantially limits the range of effects that can appear. For instance, in the spherical shell simulation the combination of stable stratification, incompressibility and axial symmetry allows only for toroidal plastic flow velocities, whereas a fully three dimensional model would allow also meridional flows. Even under such constraints however, the inclusion of plastic flows is an additional piece of physics that is important for the crust evolution. Furthermore, while plastic flows are obviously important for strongly magnetised neutron stars, the outer layers of even moderately magnetised ones may be impacted by this effect especially near their surface, and affect not only the magnetar population but also normal pulsars. 

\section*{Acknowledgements}
The author thanks the organisers of the conference for the invitation for the talk. This work has been supported financially  from Grant FK 81641 ``Theoretical and ComputationalAstrophysics'', ELKE - University of Patras.

\bibliographystyle{iaulike}

\bibliography{Bibtex1}

\end{document}